\documentclass[conference]{IEEEtran}
\IEEEoverridecommandlockouts

\def\BibTeX{{\rm B\kern-.05em{\sc i\kern-.025em b}\kern-.08em
    T\kern-.1667em\lower.7ex\hbox{E}\kern-.125emX}}
\usepackage{amsmath,amssymb,amsfonts}
\usepackage{algorithmic}
\usepackage{graphicx}
\usepackage{textcomp}
\usepackage{xcolor}
\usepackage{mathtools}
\usepackage{multirow}
\usepackage{subcaption}
\usepackage{xcolor}
\usepackage[linesnumbered,ruled,vlined]{algorithm2e}
\usepackage{setspace}
\usepackage{import}
\usepackage{lipsum}
\usepackage[outdir=./figures/]{epstopdf}
\usepackage{comment}
\usepackage{bbm}
\usepackage{authblk}
\newcommand*{\affaddr}[1]{#1} 
\newcommand*{\affmark}[1][*]{\textsuperscript{#1}}
\newcommand*{\email}[1]{#1}

\begin{document}
\title{Physical-Layer Authentication of Commodity Wi-Fi Devices via Micro-Signals on CSI Curves}

\author{
Ruiqi Kong\affmark[1] and He (Henry) Chen\affmark[1,~2]\\
\affaddr{\affmark[1]Department of Information Engineering, The Chinese University of Hong Kong, Hong Kong SAR, China}\\
\affaddr{\affmark[2]Shun Hing Institute of Advanced Engineering, The Chinese University of Hong Kong, Hong Kong SAR, China}\\
\email{E-mail: \{kr020, he.chen\}@ie.cuhk.edu.hk}\\
\thanks{This research was supported in part by project \#MMT 79/22 of Shun Hing Institute of Advanced Engineering, The Chinese University of Hong Kong. The authors would like to thank Soung Chang Liew for his insightful discussions on LS-based sparse channel estimation.}
\vspace{-3em}}

\maketitle

\begin{abstract}
This paper presents a new radiometric fingerprint that is revealed by micro-signals in the channel state information (CSI) curves extracted from commodity Wi-Fi devices. We refer to this new fingerprint as ``micro-CSI''. Our experiments show that micro-CSI is likely to be caused by imperfections in the radio-frequency circuitry and is present in Wi-Fi 4/5/6 network interface cards (NICs). We conducted further experiments to determine the most effective CSI collection configuration to stabilize micro-CSI. To extract micro-CSI from varying CSI curves, we developed a signal space-based extraction algorithm that effectively separates distortions caused by wireless channels and hardware imperfections under line-of-sight (LoS) scenarios. Finally, we implemented a micro-CSI-based device authentication algorithm that uses the k-Nearest Neighbors (KNN) method to identify 11 COTS Wi-Fi NICs from the same manufacturer in typical indoor environments. Our experimental results demonstrate that the micro-CSI-based authentication algorithm can achieve an average attack detection rate of over 99\% with a false alarm rate of 0\%.

\end{abstract}
\begin{IEEEkeywords}
Physical layer authentication, radiometric fingerprinting, channel state information, commodity Wi-Fi.
\end{IEEEkeywords}

\section{Introduction}

Radiometric fingerprinting techniques have been proposed to complement conventional cryptography-based authentication mechanisms for boosting wireless network security. Radiometric fingerprinting techniques are appealing as they leverage the inherent and unique characteristics of radio-frequency (RF) circuitry imperfections of wireless transmitters, which are hard to impersonate for attackers and easy to maintain for legitimate devices \cite{zhang}. More specifically, RF circuitry imperfections are embedded in all emitted signals and manifest as distortions that deviate from standard signals. Such signal distortions are negligible to the decoding of conveyed information, but can be distinguishable for different devices so that they can be leveraged for device authentication. Furthermore, by using RF circuitry’s inherent properties, radiometric fingerprinting can save computational resources needed in cryptography-based mechanisms for key generation and management, making it appealing for low-cost IoT devices. 

Thanks to its untamable properties, radiometric fingerprinting has attracted enormous interest and achieved satisfactory performance in the context of wireless device classification and/or authentication in recent years. However, existing radiometric fingerprinting mechanisms often need to rely on physical-layer signal samples to extract fingerprints. In practice, wireless chipsets do not output physical-layer samples to higher layers for authentication purposes. As such, previous works used dedicated and expensive instruments, such as vector signal analyzers and software-defined radios, to acquire physical-layer samples. This largely hinders the practical usability of these sample-based radiometric fingerprinting in commercial off-the-shelf (COTS) systems.

Meanwhile, several channel state information (CSI) tools \cite{atheros,linux,nexmon,picoscenes} have been developed and widely used in wireless sensing and localization applications in recent years. These works demonstrated that COTS Wi-Fi devices can report CSI to higher layers for other purposes, once a CSI tool has been properly installed. We remark that in wireless communication systems, CSI needs to be estimated at receiver side for equalizing the distortions induced by wireless channels before decoding the transmitted symbols. Therefore, the acquisition of CSI at higher layers does not need extra equipment, since it has been available at the physical layer of wireless communication systems. Furthermore, since the transmitted signals go through RF circuits before being emitted to wireless channels, the CSI estimated at receiver side incorporates the distortions induced by both wireless channels and transmitter's RF circuitry imperfections.
The above two observations motivate us to explore the feasibility of extracting radiometric fingerprints from CSI measurements. 

In this work, we report a new radiometric fingerprint, which manifests itself as the micro-signals embedded on the CSI curves (versus subcarrier index) of COTS Wi-Fi devices. We term this fingerprint as micro-CSI hereafter. We conduct several experiments to confirm that micro-CSI is most likely caused by RF circuitry imperfections and exists in Wi-Fi 4/5/6 NICs. As such, micro-CSI can be used as a radiometric fingerprint for physical-layer authentication (PLA) of Wi-Fi NICs. To that end, we devise and evaluate the first-of-its-kind micro-CSI based device authentication mechanism. Specifically, we first figure out the most favorable system configuration for acquiring CSI measurements that can make micro-CSI more stable. Additionally, we have developed a new signal space-based approach to extract the micro-CSI from varying CSI measurements of LoS scenarios. We finally implement a micro-CSI-based device authentication algorithm, which adopts k-Nearest Neighbors algorithm to identify 11 COTS Wi-Fi NICs from the same manufacturer in different indoor environments. {The experimental results show that our authentication algorithm can achieve over 99$\%$ attack detection rate with 0$\%$ false alarming rate.}

\textbf{Related work}: We noticed a handful of work that also explored the possibilities of extracting radiometric fingerprints from CSI \cite{hua,liu,lin2020} of wireless connections. More specifically, the authors in \cite{hua} extracted carrier frequency offset (CFO) from CSI measurements and achieved the highest performance in terms of attack detection rate (ADR) of 97.24$\%$ with the false alarm rate (FAR) being 1.47$\%$ when conducting experiments using 8 commercial APs by using 5000 CSI measurements for each authentication. Later, the authors in \cite{liu} extracted nonlinear phase errors from CSI and achieved up to 97$\%$ classification accuracy of 30 Wi-Fi devices by using 200 CSI measurements for one fingerprint. Authors in \cite{lin2020} extracted power amplifier (PA)-induced power variance from 2000 CSI measurements to constitute one fingerprint, which can achieve an overall 93\% ADR and a 3\% FAR when authenticating 12 commercial APs. Our work complements the above works in the sense that micro-CSI can be used together with other existing fingerprints to boost the accuracy of CSI-based PLA.  

\section{Observations of Micro-CSI}\label{s2}

\begin{figure}
  \centering
  \includegraphics[width=\linewidth
  ]{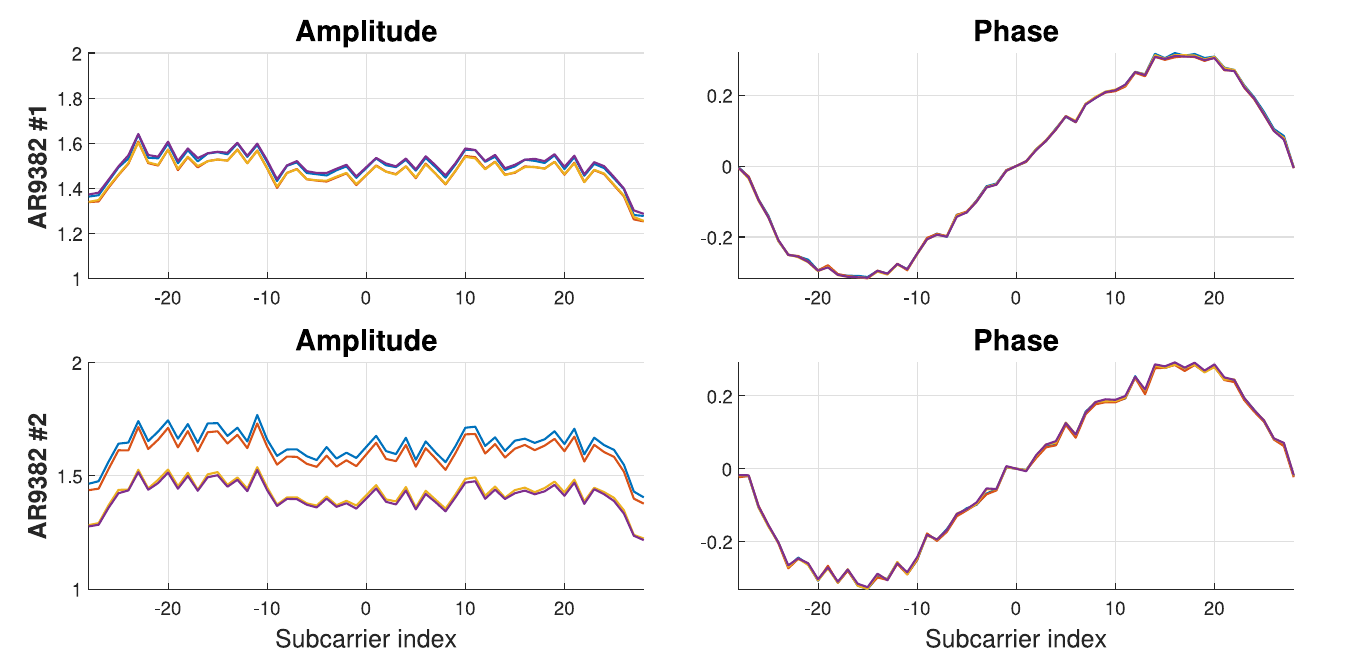}
  \caption{Similar micro-signals appear in CSI after suppressing the effects of random noises, where different colored curves represent averaged CSI of different measurement groups. The micro-signals of different NICs look quite different.}
  \label{variation}
  \vspace{-1em}
\end{figure}

A new RF distortion was observed when we used an Atheros AR9580 as the receiver to collect CSI measurements of an Atheros AR9382 NIC via the Atheros CSI tool \cite{atheros}. In our experiment, we set the system to work in a bandwidth of 20 MHz, having 56 subcarriers, and the transceiver is connected by an RF cable of 20cm and an attenuator of 30dB. The micro-signals (micro-CSI) on CSI curves look random when we observe them at a single-measurement granularity. This observation made us induce that micro-CSI could be caused by random noises. To verify whether micro-CSI originates from random noises, we tried to suppress noise impacts by averaging over multiple replicate CSI measurements of the same Wi-Fi NIC. By using the cable, we can safely assume that the channel keeps almost unchanged so that the collected CSI can be treated as replicate measurements. We started by dividing each 100 consecutive CSI measurements into one group and calculating the averaged CSI amplitude and averaged CSI phase of each group. The averaged CSI amplitude and phase of each measurement group are presented in Fig. \ref{variation}, in which each curve corresponds to one group. We noticed that the micro-CSI on the averaged CSI curves of different groups become relatively stable and exhibit highly similar patterns for the same NIC. Besides, the micro-CSI on two Atheros AR9382 NICs are different, which indicates that micro-signals are caused by RF distortions. Furthermore, as shown in Fig. \ref{variation}, the micro-CSI is distinguishable between the two NICs, even if they are of the same model that may experience similar RF manufacturing imperfections.

As the Atheros CSI tool is only available for AR93/94/95 series of Atheros NICs as the tool was developed based on the Ath9k (ar9003) driver \cite{ath9k}, we subsequently figured out whether micro-CSI is only introduced by the imperfections of the Atheros NICs. To that end, we used a set of Xilinx FPGA-based software-defined radio devices to collect CSI of different-brand NICs.  
We observed that the micro-CSI shows up on all tested Wi-Fi 4/5/6 NICs and is distinguishable between them. Due to space limitations, we cannot provide the corresponding CSI curves here. 
In summary, micro-CSI is highly related to RF distortions and becomes relatively stable after suppressing noises. 

\section{Micro-CSI as Radiometric Fingerprint}\label{s3}
This section will elaborate on how to realize PLA of COTS Wi-Fi devices using micro-CSI as the RF fingerprint. Specifically, we will discuss how to conduct signal acquisition, fingerprint extraction, and fingerprint matcher \cite{denev}.

\subsection{Signal Acquisition}\label{sa}
Signals are transmitted from unauthorized devices and contain unique signal properties (i.e., fingerprints) for device authentication. The acquisition step should retain the unique fingerprint properties that the authentication relies on, rather than influencing or degrading the fingerprints, e.g., by introducing measurement errors \cite{denev}. Under this requirement, when it comes to our CSI-based fingerprinting system, we need to figure out what is the most favorable configuration of CSI collection for precise fingerprint extraction. 

In Wi-Fi systems, CSI is estimated based on the shared knowledge of reference sequences at transmitter and receiver sides. The CSI can be estimated by comparing the received reference sequences with the transmitted version. {Recent Wi-Fi systems adopt multiple-input multiple-output (MIMO) technologies on top of OFDM, in which multiple antennas are used to enable the transmission and/or reception of multiple spatial data streams at the same time. In MIMO-OFDM systems, space-time block coding (STBC) encoder at the transmitter and decoder blocks at the receiver are used to support the concurrent transmission of multiple orthogonal spatial streams \cite{stbc}. By applying orthogonal STBC, the system can transmit multiple reference sequences simultaneously and estimate the CSI of all transceiver links at the same time. 

Different transmitter chains use separate radio frequency circuits in MIMO-OFDM systems, and thus they may have different fingerprints. Nevertheless, the fingerprints of different transmitter chains do not enjoy the orthogonality as the transmitted sequences do.} As such, the MIMO configuration may mix fingerprints of multiple transmitter chains together, making the fingerprints of different Wi-Fi NICs look more similar. In this regard, we choose to use the signal-antenna configuration at the transmitter side during the signal acquisition to collect more stable and distinct micro-CSI in varying environments. {We remark that single-antenna IoT devices have been widely deployed in smart homes and intelligent factories {\cite{al2015internet,chen2021securepilot}}. Besides, for Wi-Fi NICs that equipped with two or more RF chains, the number of “active” chains can be dynamically configured by application layer commands or modifying the NIC driver \cite{picoscenes}.} Furthermore, such single-antenna configuration\footnote{To fulfil this requirement, we can implement micro-CSI-based authentication in a challenge-response manner. Specifically, an authentication server can send a challenge packet to ask an unauthorized device to send a group of packets using the single antenna setup.} is only needed when a device needs to go through the occasional authentication process. It will not hinder the use of multiple antennas for normal data communications.

We also conducted an experiment to verify the influence of different receiver chains on micro-CSI. In this experiment, we used two NICs (AR9382) as the transceiver. Specifically, the transmitter was configured to use a single antenna and the receiver used two antennas. Our results show that despite the wireless channels between transmitter chain $\#1$ with receiver chains ($\#1,\#2$) are different, receiver chains of the same device cause insignificant differences on micro-CSI. Thus, SIMO (single-input multiple-output) configuration allows collecting CSI of multiple tranceiver links carrying the micro-CSI of the single-antenna transmitter at the same time. This means from each CSI measurement, we can extract multiple replica of the same micro-CSI, which can be used to further suppress noises. Considering this benefit, we adopted SIMO configuration in all the results presented hereafter.

\subsection{Signal Space-Based Fingerprint Extraction}

Each CSI data comprises information about the wireless channel, micro-CSI induced by hardware distortions, and noises. As we can see from Fig. \ref{variation}, wireless channel information and micro-CSI entangle in the frequency domain, making the construction of channel-independent fingerprint non-trivial. In the following, we first present the signal models of the channel information and micro-CSI before presenting a signal space-based fingerprint extraction method for strong LoS conditions.

The estimation of CSI in 802.11 protocols is accomplished based on the Long Training Symbol (LTS) included in the preamble part of each packet. Commodity Wi-Fi systems only use a subset of subcarriers for communication, denoted by $\mathcal{K}$. In the case of 802.11n, a 64-point DFT is adopted, with only 56 subcarriers being used for transmission (i.e., $|\mathcal{K}| = 56$). Denote by column vectors ${\bold{t}} = ({t}_n)_{n\in\mathcal{N}}$ and $\Tilde{\bold{t}} = (\Tilde{t}_k)_{k\in\mathcal{K}}$ the LTS in time domain and frequency domain, respectively, where $\mathcal{N} = \{1,\dots,N\}$ with $N$ being the DFT length. The transmitter's introduced distortions (i.e., micro-CSI) in the time and frequency domains are respectively represented by $\bold{d} = (d_n)_{n\in\mathcal{N}}$ and $\Tilde{\bold{d}} = (\Tilde{d}_k)_{k\in\mathcal{K}}$. We further have $\Tilde{\bold{t}} = \bold{F}_{\mathcal{K},\mathcal{N}}\bold{t}$ and $\Tilde{\bold{d}} = \bold{F}_{\mathcal{K},\mathcal{N}}\bold{d}$, where $\bold{F}$ is the full unitary DFT matrix, and $\bold{F}_{\mathcal{K},\mathcal{N}}$ is the sub-matrix of $\bold{F}$, comprising all rows with indexes in $\mathcal{K}$ and all columns with indexes in $\mathcal{N}$. 
{Mathematically, we can regard the hardware distortions as some deviations made to the standard LTS samples. Under this model, the time-domain signal emitted to the air can then be written as $\bold{s=t+d}$.}
After the cyclic prefix (CP) cut, the received signal can be written as
$\bold{y=h \ast s + z}$, where $\ast$ denotes circular convolution, $\bold{h}$ is the discrete-time equivalent channel, and $\bold{z}$ is complex white Gaussian noises. 
After synchronizing the received samples, the CSI in the frequency domain can be estimated by least squares (LS) estimation \cite{book80211}. Specifically,  
\begin{equation}\label{est_CSI}\Tilde{\bold{c}}=\Tilde{\bold{y}}\circ\Tilde{\bold{t}}=\Tilde{\bold{h}}\circ(\bold{1}+\Tilde{\bold{d}}\circ\Tilde{\bold{t}})+\Tilde{\bold{z}} = \Tilde{\bold{h}}+\Tilde{\bold{h}}\circ\Tilde{\bold{d}}\circ\Tilde{\bold{t}}+\Tilde{\bold{z}} ,
\end{equation}
where $\circ$ is element-wise multiplication and $\bold{1}$ represents the column vector with all 1's, and $\Tilde{\bold{z}} = (\Tilde{z}_k)_{k\in\mathcal{K}}$ is the frequency-domain noise vector with  $\Tilde{z}_k\sim\mathcal{N}(0,\sigma^2)$. We can see from (\ref{est_CSI}) that the channel information $\Tilde{\bold h}$ and the hardware distortion $\Tilde{\bold d}$ entangle together. To circumvent the problem, our basic idea is to analyze the CSI measurements from the signal space perspective. Specifically, thanks to the inherent channel sparsity, the number of taps of wireless channels in time domain (i.e., $\bold{h}$) is often much smaller than the DFT length $N$. As such, the micro-CSI that resides in other unoccupied dimensions can be separated from the channel information. As the first attempt, we start with the strong LoS scenario.

Under strong LoS, we can safely assume that the first arrived physical path has the strongest power and dominates the channel response. However, a single physical path does not mean that the time-domain CSI $\bold h$ only has one tap. This is because a single path can lead to multiple channel taps when the propagation delay is not an integer multiple of the sampling rate $T_s$, which is known as the shaping leakage \cite{bala2013shaping}. The number of leakage taps depends on the pulse-shaping filter, accounting for leakage on both earlier and later taps. In 802.11 protocols, frame synchronization is achieved using correlation-based algorithms that align with the strongest signal tap, which is also the central tap. Besides, taps before the strongest tap will be circularly shifted by $N$. Therefore, the set of taps that carries channel information is $\mathcal{L} =\{-N_p,..., N_p\} \bmod N$, where $N_p$ is the number of leaked taps on each side. An example of a single-path leaked channel is given in \cite[Fig. 1]{xk_spawc}. Due to the rapid decay of the impulse response of the pulse shaping filter, we set  $N_p=8$ in all our experiments. The underlying assumption is that the pulse shape will decay to zero after 8 sampling time intervals.

\begin{figure}
    \centering
\includegraphics[width=\linewidth]{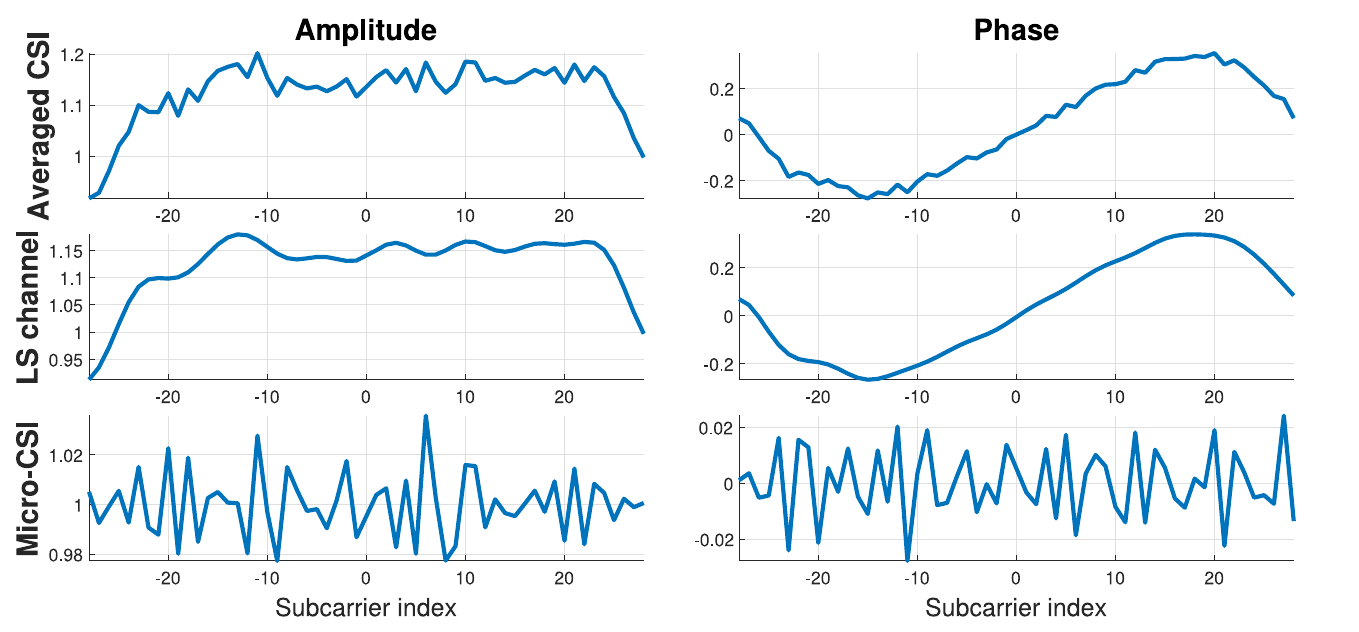}
  \caption{Extracted Micro-CSI with $N_{csi}=100$.}
  \label{fingerprint}
   \vspace{-2em}
\end{figure}

As discussed in Section \ref{s2}, the micro-CSI, due to its small scale, can easily be contaminated by noises. To suppress the noises and other system errors, we take an average of a number of consecutive CSI measurements before extracting the micro-CSI. Since the CSI collection time is short, it is reasonable to assume that the channel remains constant during a batch of CSI measurements. Hereafter, $N_{csi}$ represents the number of consecutive CSI measurements used to construct one micro-CSI. Later in Section \ref{s4}, we will evaluate how the value of $N_{csi}$ affects the accuracy of the device authentication performance. 
The sample mean of $M$ consecutive  observations of the estimated CSI is
\begin{align}
\bold{\tilde{c}}_e &= \frac{1}{M}\sum_{m=1}^M\bold{\tilde{c}}_m
                       \approx \Tilde{\bold{h}}+\Tilde{\bold{h}}\circ\bold{\tilde{f}}+\bold{\tilde{z}}_e,
\end{align}      
where $\Tilde{\bold{f}}=\Tilde{\bold{d}}\circ\Tilde{\bold{t}}$ can be treated as the fingerprint, $\bold{\tilde{z}}_e = (\bar{\tilde{z}}_k)_{k\in\mathcal{K}}$ with $\bar{\tilde{z}}_k \sim \mathcal{N}\left(0,\frac{\sigma^2}{M}\right).$ 
To construct a channel-independent fingerprint, we need to eliminate the channel $\Tilde{\bold{h}}$ from the averaged CSI $\bold{\tilde{c}}_e$. Under strong LoS, we can assume that each element of the term $\Tilde{\bold{h}}$ is much larger than that of the term $\Tilde{\bold{h}}\circ\bold{\tilde{f}}$, considering the small scale of $\bold{\tilde{f}}$. This indicates that the values on these time-domain taps occupied by the channel can be considered to be solely contributed by the channel information. In this context, leveraging the fact that the number of channel taps is limited, we can apply the LS method \cite{ls} to estimate the frequency-domain channel information by
\begin{align}
    \hat{\Tilde{\bold{h}}}_e&=\bold{F}_{\mathcal{K},\mathcal{L}}(\bold{F}_{\mathcal{K},\mathcal{L}}^H\bold{F}_{\mathcal{K},\mathcal{L}})^{-1}\bold{F}_{\mathcal{K},\mathcal{L}}^H\Tilde{\bold{c}}_e \approx \Tilde{\bold{h}},
\end{align}
where $(\cdot)^{H}$ and $(\cdot)^{-1}$ represent the Hermitian transpose and inverse operations, respectively. The first step $\bold{F}_{\mathcal{K},\mathcal{L}}^H\Tilde{\bold{c}}_e$ transform averaged CSI to time domain and keep only the channel taps within $\mathcal{L}$. The second step $(\bold{F}_{\mathcal{K},\mathcal{L}}^H\bold{F}_{\mathcal{K},\mathcal{L}})^{-1}$ is a correct term because $\bold{F}_{\mathcal{K},\mathcal{L}}$ is a partial DFT. The third step $\bold{F}_{\mathcal{K},\mathcal{L}}$ transforms back to the frequency domain. The channel-independent fingerprint can be estimated by element-wise dividing $\Tilde{\bold{c}}_e$ by $\hat{\Tilde{\bold{h}}}_e$. That is, {$\hat{\Tilde{\bold{f}}}= \Tilde{\bold{c}}_e ./
\hat{\Tilde{\bold{h}}}_e\approx \bold{1}+\bold{\tilde{f}}$.} Fig.~\ref{fingerprint} shows the micro-CSI extracted from one averaged CSI measurement, where $N_{csi}=100$. We can see that our algorithm can effectively extract the micro-CSI embedded on CSI.

\subsection{{Fingerprint Matcher}}

The fingerprint matcher is the process of matching unauthorized device with legitimate devices by comparing their fingerprints, where the fingerprints of legitimate devices are pre-collected and stored in a library. In this work, the detection of the rogue device is performed by the k-Nearest Neighbors (KNN) anomaly detection algorithm, which is simple yet widely used \cite{knn}. The fundamental principle of KNN is that the observations of the same cluster are expected to be in proximity to each other, and outliers should be further away from the cluster of similar observations. The KNN algorithm detects outliers by leveraging the relationship between data points of neighborhoods. The further a data point is from its neighbors, the more likely it is an outlier. 

In our subsequent experiments, each transmitting NIC was configured in single-antenna mode, while the AP was equipped with two antennas. As discussed in Section \ref{sa}, receiver chains {of the same device} cause insignificant differences on micro-CSI, therefore micro-CSIs from different receiver chains can thus be used as if they were collected by the same antenna to suppress noises. Then, we extract micro-CSI from a group of $N_{csi}$ CSI measurements.
After that, we calculate and average the Euclidean distances between micro-CSI and its corresponding $K$ nearest neighbors of claimed identity stored in the fingerprint library. We thus obtain the distance value. Finally, a threshold is used to determine whether the unauthorized device is the claimed legitimate device or not. Specifically, when the distance is higher than the given threshold, the device is reported as a rogue device, and the authentication fails. By contrast, the device is considered to be the claimed device when the distance is below the threshold.

\section{Performance Evaluation}\label{s4}
In this section, we present the device authentication performance of micro-CSI in two typical indoor LoS environments.

\subsection{Experiment Setup}
\textbf{Device Configurations}: We used 11 Wi-Fi NICs randomly selected from three series of Atheros NICs for this performance evaluation experiment, including 1 AR9462, 1 AR9485, 1 AR9565, 1 AR9580, 1 AR9380, 6 AR9382. We set up a mini-PC equipped with an AR9580 NIC and two antennas as the AP to collect CSI measurements from 11 Atheros NICs using the Atheros CSI Tool \cite{atheros}. The AP was configured to work on the 2.4GHz Channel 10 with a bandwidth of 20 MHz (i.e., the center frequency is 2457 MHz) and run the 802.11n protocol. Every unauthorized NIC was installed on another mini-PC and connected to the AP as a station, where only one device is connected to the AP at any time. All the NICs were equipped with a single antenna and emitted one packet approximately every 50 microseconds.

\textbf{Data Collection}: We conducted CSI collections for each NIC in Room A (research office) and Room B (common room), respectively. In both rooms, the transceivers were placed several meters away from each other in LoS conditions. 
We conducted experiments in two rooms on different days to make the collected CSI space in time and environments for better evaluating the robustness of micro-CSI. The AP acquired $2 \text{ (Rooms)}\times60000\text{ (Packets)}=120000$ CSI measurements for each NIC. 

\begin{table}
  \caption{Attack Detection Rate.}
 \begin{subtable}[h]{\linewidth}
  \label{a1}
  \centering
   \small
  \begin{tabular}{l|l|l|l|l|l}
    \hline
    $N_{csi}\rightarrow$&10&20&50&100&200\\
    \hline
    \multirow{1}{5em}{FAR= 0\%}
    &16.35\%&48.08\%& 80.41\% & 96.82\%&99.21\%\\
    \hline
      \multirow{1}{5em}{FAR$\leq$ 3\%}
    &93.18\%&96.16\%& 98.68\% & 99.48\%&99.77\%\\
  \hline
\end{tabular}
\vspace{1mm}
\end{subtable}

\label{adr}
\vspace{-1.5em}
\end{table}

\subsection{{Authentication Performance}} 
We adopted the attack detection rate (ADR) and false alarm rate (FAR) as performance metrics, where the ADR is the probability of successfully detecting a rogue device, and the FAR is the probability of mistakenly rejecting a legitimate device. In all the results presented in this section, we used $S=60000/N_{csi}$ fingerprints collected in Room A to construct the fingerprint library of each legitimate device. To challenge our framework, we used the fingerprints of devices collected in Room B as test fingerprints. This unfriendly setup evaluates the capabilities of micro-CSI in authenticating devices under different environments. Besides, we set $K$ to be $\sqrt{S}$ in our KNN algorithm. That is, we calculate and average the Euclidean distances between each test fingerprint and its $K$ nearest neighbors of claimed identity in the fingerprint library.

\begin{figure}
  \centering
  \includegraphics[width=\linewidth
  ]{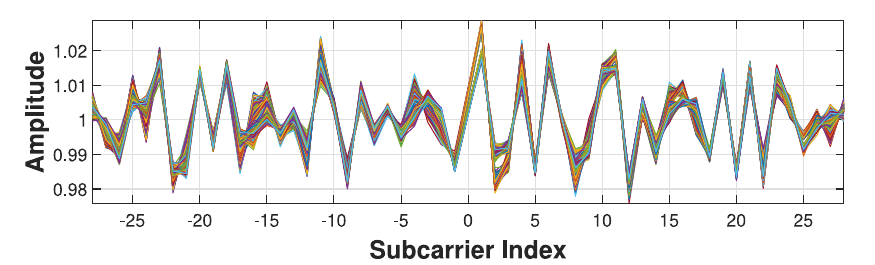}
  \caption{Stability of fingerprints, when $N_{csi}=200$. Each colored curve represents one fingerprint.}
  \label{sc}
  \vspace{-2em}
\end{figure}

In our evaluation, we rotated each of the 11 NICs to serve as the legitimate device, which was attacked by the remaining 10 NICs, one at a time. The number of test fingerprints from each attacker is equal to the number of test fingerprints from the legitimate NIC. Table \ref{adr} shows the averaged ADR value of NICs to be attacked by the remaining NICs.
As our data spacing in time and environments, the experimental results show that micro-CSI is robust to time and environmental changes in LoS conditions. Furthermore, as the number of CSI measurements used for constructing one fingerprint (i.e., $N_{csi}$) increases, our scheme can achieve 99.21$\%$ ADR with 0$\%$ FAR even when rogue and legitimate NICs are of the same brand. 
The worst performance in our evaluation happened when an AR9382 NIC acted as the legitimate NIC. Specifically, it achieved 100\% ADR with 0$\%$ FAR when subjected to attacks from NICs of different models. However, when attacked by NICs of the same model, it only achieved an average of 84.67\% ADR with 0$\%$ FAR. This discrepancy in performance can be attributed to two primary factors. Firstly, there is a higher similarity between fingerprints obtained from NICs of the same model, resulting in smaller inter-cluster distances. Secondly, the legitimate fingerprints of certain subcarriers display variations, leading to larger intra-cluster distances. These variations exacerbate the difficulty in accurately detecting attacks from the same model. Fig.~\ref{sc} shows the varying stability of fingerprints across different subcarriers. This motivates us to adopt subcarrier selection to further boost performance in our future work. Last, we remark that even when $N_{csi}=200$, the time required for collecting the CSI measurements only takes around 10 milliseconds, showing that our scheme can achieve fast authentication.

\section{Conclusions} \label{s5}
In this paper, we reported a new radiometric fingerprint embedded on channel state information (CSI) curves, named micro-CSI. Our experiments confirm that micro-CSI is most likely caused by RF circuitry imperfections and exists in Wi-Fi 4/5/6 NICs. Additionally, we have developed a signal space-based framework to extract the micro-CSI from varying CSI measurements under LoS environments. Furthermore, we designed and implemented a micro-CSI-based device authentication scheme, which adopts the simple k-Nearest Neighbors (KNN) algorithm to identify 11 COTS Wi-Fi devices from the same manufacturer in different indoor environments. Our experimental results showed that despite the use of the simple KNN algorithm, our scheme can achieve over 99$\%$ attack detection rate with 0$\%$ false alarming rate. 
Future works will include the implementation of subcarrier selection of fingerprints and further evaluation of the performance of micro-CSI-based authentication under non-line-of-sight conditions, where advanced algorithms, such as deep learning-based approaches, can be explored to further improve the performance.
\vspace{-1mm}
\bibliography{microcsi_spawc} 
\bibliographystyle{ieeetr}

\end{document}